\documentclass[10pt,twocolumn]{IEEEtran}
\usepackage{amsmath,amsthm,amssymb,epsfig,subfigure,cite}
\usepackage{algorithm,algpseudocode}

\newtheorem{theorem}{\hspace*{1pc}Theorem}
\newtheorem{corollary}{\hspace*{1pc}Corollary}
\newtheorem{lemma}{\hspace*{1pc}Lemma}

\begin{document}

\bibliographystyle{unsrt}

\setlength{\parindent}{1pc}

\title{Fitting a Model to Data in Loss Tomography}
\author{Weiping~Zhu,\IEEEmembership{ member, IEEE}  \thanks{Weiping Zhu is with University of New South Wales, Australia}}
\date{}
\maketitle

\begin{abstract}
 Loss tomography has received considerable attention in
recent years and a number of estimators have been proposed. Although
most of the estimators claim to be the maximum likelihood estimators,
the claim is only partially  true since  the maximum likelihood
estimate can be obtained at most for a class of data sets.
 Unfortunately, few people are aware of this restriction that leads to a misconception that an estimator
 is applicable to
 all
data sets as far as it returns a unique solution.  To correct this, we
in this paper point out the risk of this misconception and illustrate
the inconsistency between data and model in the most influential
estimators.  To ensure the model used in estimation consistent with
the data collected from an experiment,  the data sets used in
estimation are divided into 4 classes according to the characteristics
of
 observations. Based on the classification, the validity of an
 estimator is defined and the validity of the most influential estimators
 is
 evaluated. In addition, a number of
 estimators are proposed, one for a class of data sets that have been overlooked. Further,
 a general estimator is proposed that is applicable to all data
 classes. The discussion starts from the tree topology and end at the general topology.
\end{abstract}


\section{Introduction}

Network measurement has received considerable attention in recent
years since it can not only provide necessary information for network
modeling but also verify some of the assumptions made on ad hoc basis
for network modeling. In contrast to direct measurement that is only
suitable for a small network, network tomography is
 proposed for a large network \cite{YV96}. Network tomography, as a methodology, differs from
direct measurement in many ways, and the most important one is the use
of statistical inference to accomplish the task that involves probing,
modeling, and estimation. As a new methodology,
 a large
number of works have been carried out in recent years and the results
reported so far cover loss tomography \cite{CDHT99},
\cite{Duffield2002}, \cite{ZG05}, \cite{ZD09}, \cite{Zhu09}, delay
tomography \cite{LY03}, \cite{TCN03}, \cite{PDHT02}, \cite{SH03},
\cite{LGN06}, loss pattern tomography \cite{ADV07}, source-destination
traffic matrix \cite{LY06}, and shared congestion flows \cite{RKT02}.
Despite the overwhelming enthusiasm and a wealth of publications in
this area, network tomography is in its infancy and there is more work
that needs to be done. In addition, some of the characteristics that
seems to be fully investigated still require further studies. For
instance, although loss tomography has been studied for more than 10
years and the likelihood equation of the tree topology has been
available for over 10 years \cite{CDHT99}, few  are aware that the
likelihood equation is restricted to a specific class of data. That is
also true to almost all other estimators proposed in the past. Using
those estimators without checking data type may result in unexpected
consequence. This paper is devoted to address the validity of an
estimator in the context of data classes and propose estimators for
those unidentified data classes.

The fundamental principle of loss tomography is built on statistical
inference, where parametric estimation is frequently used to fit a
statistical model to data (observation), and then the maximum
likelihood estimation is used to find the unknown parameters of the
model. Unfortunately, the principle has been partially overlooked by
almost all of the previous works. In addition, the few exceptions,
such as \cite{CDHT99} and \cite{Zhu09}, that consider this issue take
a different approach to handle this. Rather than fitting a model to
data, the exceptions go in the opposite direction and try to fit data
to a predefined model. If a data set does not fit the model, the data
set is either discarded or skipped. Also, the discussions presented in
\cite{CDHT99}, \cite{Zhu09} are far from complete, which fail to
identify all unsuitable data sets and, of course, fail to propose
alternative likelihood equations for the unidentified data sets.
Furthermore, there has been no discussion of the validity of an
estimator (a likelihood equation) in regard to data sets. Without
knowing those, an estimator may be mistakenly used on a data set to
that it does not fit and returns an incorrect estimate of the unknown
parameter(s). The incorrect estimate can be very different from the
correct one that invalidates the estimate from being used in traffic
control or modeling. Apart from presenting the problems of previous
works, we also provide solutions, where a number of estimators, one
for a data class, are proposed in this paper to overcome the problems.
More, a general estimator applicable to all type of data is proposed
in this paper. More over, the discussion covers both tree and general
topologies.

Previous works on loss tomography were focused on proposing estimators
and proving there is a unique solution from the proposed estimator. If
an estimator cannot return a unique solution or cannot return a
solution, the data set used in the estimation is considered
inconsistent with the estimator and skipped or discarded \cite{CDHT99}
 \cite{ZD09}. Although it is fundamentally important in terms of identifiability to make sure
 a unique solution from a likelihood equation, we must know that the likelihood equation is from
a likelihood function and the likelihood function is from a data set.
Only if the likelihood equation used in estimation fits the nature of
the data set, is the unique solution obtained from the equation the
maximum likelihood estimate (MLE), if the maximum likelihood principle
is used in the estimation.  To remedy this, data obtained from
experiments are divided into 4 exclusive classes according to the
characteristics of intersection and partition in observation. The 4
classes are called perfect, chained-only, partitioned, and chain
partitioned, respectively. Each class requires a likelihood equation
and the likelihood equations proposed in \cite{CDHT99}
 \cite{ZD09} only suit to the data sets
of the perfect class. Although the perfect class is the most likely
scenario among the 4 classes, in particular when the number of probes
sent to receivers approaches infinite, i.e. $n \rightarrow\infty$,
other scenarios do occur from time to time if $n < \infty$.

\subsection{Contribution and Paper Organization}

 Data consistency raised in
\cite{CDHT99} aims to eliminate 3 types of data from estimation since
the likelihood equation proposed in that paper cannot find a solution
for the 3 types of data. Nevertheless, \cite{CDHT99} falls short of
considering whether a unique solution returned by the likelihood
equation is the MLE. In other words, the unique solution from a
likelihood equation is only the necessary condition of the MLE, while
the sufficient condition requires the likelihood equation fits the
data set. To make this possible,  we in this paper present the
relationship between data and likelihood equations, and emphasize that
MLE can only be obtained if and only if a likelihood equation fits to
the data and there is a unique solution to the likelihood equation.
The contribution of this paper can be divided into two parts: identify
problems and find solutions, which are detailed as follows:

\begin{enumerate}
\item To improve our understanding of statistical inference in loss
tomography, this paper reiterates the importance of the statistical
principle of fitting a model to data in the context of loss
tomography. It points out that  an estimator is only valid if the
model used by the estimator fits the data collected from observation.
It further examines the validity of the most influential estimators
proposed for loss tomography and identifies the pitfalls of the
estimators.
\item To solve the problems, data used in estimation are divided into 4 classes on the basis of intersection and partition in
the observation of descendants. The estimators proposed so far only
fit to one of the 4 classes. Then, a number of estimators are proposed
for the other 3 classes that have been overlooked, including a general
estimator that is able to handle all data classes.
\end{enumerate}

The rest of the paper is organized as follows. In Section 2 we present
the essential background, including the notations and statistics used
in this paper. In Section 3, we present the problems existed in the
most influential estimators in details. Section 4 provides the
solutions to the problems presented in Section 3 for the tree
topology. Section 5 extends the solutions obtained from the tree
topology to the general topology. The last section is devoted to
concluding remark.

\section{Notations and Related Works} \label{section2}
The two most influential works in loss tomography, one for the tree
topology \cite{CDHT99} and the other for the general topology
\cite{ZD09}, are introduced in this section, where the latter is
developed on top of the former. Because of the relation, both have the
same restriction to the data used in estimation.

\subsection{Notation}\label{treenotation}
To assist the following discussion, the symbols used in this paper and
their definitions are introduced briefly  in this section. For those
who wants to know the details, please refer to \cite{ZD09}.

 To
collect information from a large network, a number of probes are
multicast from a number of sources located on one side of the network
to a number of receivers located on the other side of the network. The
paths from sources to receivers cover the links of interest. If there
is only a single source, the paths from the source to receiver forms a
special tree, called multicast tree, where the root only has a child.
Let $T=(V, E, \theta)$ donate the multicast tree, where $V=\{v_0, v_1,
... v_m\}$ is a set of nodes representing routers and switches of a
network; $E=\{e_1,..., e_m\}$ is a set of directed links connecting
the nodes of $V$; and $\theta=\{\theta_1,..., \theta_m\}$ is an
m-element vector, one for a link to describe the loss rate of the
link.  $R$ is used to denote all receivers.
 As a hierarchical structure, each
node in a tree except the root has a parent. Each node except leaf
ones has a number of descendants. Let $d_i$ denote the descendants of
node $i$ and $|d_i|$ denote the number of descendants in $d_i$.
Further, each multicast subtree is named by the number assigned to the
child node of the root, where $T(i)=\{V(i), E(i), \theta(i)\}, i \in
\{1,\cdot\cdot, m\}$ denotes the multicast subtree rooted at node
$f(i)$, where $V(i)$, $E(i)$ and $\theta(i)$ are the nodes, links and
parameters of the subtree. Note that a multicast subtree is different
from an ordinary one, where multicast subtree $i$ is rooted at node
$f(i)$ that uses link $i$ to connect subtree $i$. The group of
receivers attached to $T(i)$ is denoted by $R(i)$. If $n$ probes are
dispatched from the source, each probe $i=1,...., n$ gives rise of an
independent realization $X^{(i)}$ of the loss process $X$, $X_k^i=1,
k\in E$ if probe $i$ passes link $k$; otherwise $X_k^i=0$. The
observation of $\Omega=(\Omega_k), k \in V \setminus 0$ and
$\Omega_k=(X_j^{(i)})^{i=1,2,...,n}, j \in R(k)$ comprise the data set
for inference. Therefore, observations are also called data or data
sets in the following discussion.

To estimate the loss rates of a tree topology, a set of sufficient
statistics is introduced in \cite{ZD09} one for a node to denote the
number of probes reaching node $i$ confirmed from observation of
$R(i)$, $n_i(1), i \in V$.  In addition, let $n_{ij}(1), i,j \in d_k $
denote the number of probes that are observed simultaneously by at
least one receiver attached to subtree $i$ and at least one receiver
attached to subtree $j$. Similarly, $n_{ijl}(1), i, j, l \in d_k$
denotes the number of probes observed simultaneously by the receivers
attached to subtrees $i, j$, and $l$. Furthermore, we can have
statistics to count the number of probes observed simultaneously by
more descendants. This process continues until all descendants are
included, i.e., $n_{d_k}(1)$.

Given $d_k$,  we have a $\sigma$-algebra $M$ on $d_k$, and a measure
$I$ on $M$. Then, a measurable space, $(d_k, M, I)$, is established
for each node to obtain the statistics used in estimation and to
divide observations into classes, where $I(x)=\sum_{i \in {1,..,n}}
\bigwedge_{j \in x} X_j^{\{i\}}, x \in M$ counts the number of probes
observed
 simultaneously by the members of $x$. Note that $n_k(1)=I(x), \#(x)=1
 \bigwedge x \in M\setminus \emptyset$.

In contrast to the tree topology, there are multiple intersected trees
in a general network. Then, the nodes located in a shared area, called
shared segment, can observe probes sent by multiple sources. To
accommodate multiple sources, the notations defined above need to be
extended to consider the sources. Therefore, an extra symbol in most
cases is added to the corresponding notations defined above to
represent the source.  For instance, $n_i(s,1)$ denotes the number of
probes sent by source $s$ passing link $i$, $\Omega_k(s)$ denotes the
observation of $R(k)$ for the probes sent by source $s$, where $n^s$
is the number of probes sent by $s$. Note that link instead of node is
used as the reference in the general topology since there is no longer
1-to-1 mapping between nodes and links in the general topology.

\subsection{Related Works}

Multicast Inference of Network Characters (MINC) is the pioneer of
using multicast probes to create correlated observations at the
receivers of the tree topology \cite{CDHT99}, \cite{CDMT99},
\cite{CDMT99a}, where a Bernoulli model is used to describe the loss
process of a link. Using this model, the authors of \cite{CDHT99}
derive a direct expression of the pass rate of a path connecting the
source to an internal node as follows:

\begin{equation}
1-\dfrac{\gamma_k}{A_k}=\prod_{j \in d_k}(1-\dfrac{\gamma_k}{A_k}
\cdot \alpha_j) \label{mgeneral}.
\end{equation}
Using empirical probability $\hat \gamma_k=\dfrac{n_k(1)}{n}$ and
$\hat \alpha_j=\dfrac{n_j(1)}{n_k(1)}$ to replace $\gamma_k$ and
$\alpha_j$, (\ref{mgeneral}) becomes a single-variable polynomial that
has $|d_k|-1$ roots according to the fundamental theorem of algebra.
However,  we are only interested in the roots falling into the support
of $A_k$, i.e.$(0,1)$. The lemma 1 of \cite{CDHT99} proves there is a
unique solution to (\ref{mgeneral}) in the support of $A_k$ if
$\sum_{j \in d_k} \alpha_j >0$. In addition, three extreme cases are
identified and ruled out from estimation since there is no solution to
the likelihood equation if the data set used for estimation falls into
the three cases.

 Recently, Zhu proposes an analytical
solution to the general topology \cite{Zhu09}, where the likelihood
equation is as follows:
\begin{eqnarray}
 1-\dfrac{\gamma_i(k)}{A(k,i)} =
 \prod_{j \in d_i}(1-\dfrac{\gamma_i(k)\sum_{s \in S(i)} n_j(s,
1)}{A(k, i)\cdot \sum_{s \in S(i)} n_i(s,1)}) \label{generalpoly}
\end{eqnarray}
where $S(i)$ is the set of sources sending probes to node $i$ and $k$
is one of the sources. It is also proved that (\ref{mgeneral}) is a
special case of (\ref{generalpoly}). As \cite{CDHT99}, \cite{Zhu09}
also discusses the data consistent problem in the same line as its
predecessor for the general topology. In the paper, Zhu points out the
difference between the tree topology and the general one in terms of
data consistency. Despite this, the paper as its predecessor treats
the data falling into the 3 types as exceptions and eliminate them
from estimation.

Unfortunately, both, \cite{CDHT99} and \cite{Zhu09},  fail to go a
step further to consider whether the unique solution returned from
(\ref{mgeneral}) or (\ref{generalpoly}) for a data set is always the
MLE.

\section{Problem Formulation}
As stated, previous works fail to consider the impact of data on
likelihood equations. To illustrate the impact, we examine the
likelihood equations proposed in \cite{CDHT99} and \cite{Zhu09} with
imperfect data to calibrate their validity in this section.
\subsection{Statistical Implication}

Given observation $\Omega$, a likelihood function $P(\Omega|\Theta)$
is constructed as a probability measurement where $\Theta$ is a
variable. The maximum likelihood principle proposed by Fisher aims to
find the $\Theta$ that can maximize $P(\Omega|\Theta)$. The structure
of the likelihood function depends on $\Omega$, so does the likelihood
equation since it is derived from the likelihood function. Thus, the
likelihood equation as a statistical model connects some random
variables to the others and expresses the relation between the
variables. The relation can be analyzed on the basis of matching a
model to  data. Taking (\ref{mgeneral}) as an example, the both sides
of the equation denotes the loss rate of subtree $k$, where the left
hand side (LHS) uses the data obtained from observation directly to
express the loss rate of subtree $k$ while the right hand side (RHS)
uses the probability reasoning to achieve the same. The LHS can be
viewed as the data and the RHS as the model. The correspondence
between data and model becomes obvious if we expand the both sides of
(\ref{mgeneral}), where the LHS is as follows:
\begin{eqnarray}
1-\dfrac{\hat\gamma_k}{A_k}&=&1-\dfrac{n_k(1)}{n\cdot A_k} \nonumber \\
&=&1-\dfrac{1}{n\cdot A_k}\big [\sum_{j \in d_k} n_j(1)+\sum_{i,j \in
d_k} n_{ij}(1)- \nonumber \\ && \sum_{i,j,l \in d_k} n_{ijl}(1)
\cdot\cdot +(-1)^{|d_k|-1} n_{d_k}(1) \big] \label{nkexpansion}
\end{eqnarray} which is constructed from $\Omega_k$. It is easy to prove
\begin{equation}
n_k(1)\leq \sum_{j \in d_k} n_j(1)
\end{equation} since $n_{i..j}(1) \geq 0$ and the values of the terms on the RHS of (\ref{nkexpansion}) monotonically
decrease  from left to right, i.e. a term in a left summation is
larger than a term in a right summation whose subscript has one more
number than its left correspondent. Thus, $\sum_{j \in d_k}
\hat\alpha_j\geq 1$.

In contrast to the LHS, the RHS of (\ref{mgeneral}) is built on the
frequentist view that expresses the loss rate of subtree $k$ by the
product of the loss rates of the subtrees rooted at node $k$.
Expanding the RHS, we have
\begin{equation}
1-\sum_{j \in d_k}\dfrac{\gamma_j}{A_k} + \sum_{\substack{ j<k\\
j, k \in d_k}}\dfrac{\gamma_j\gamma_k}{A_k^2} \cdot\cdot +
(-1)^{|d_k|}\dfrac{\prod_{j \in d_k}\gamma_j}{A_k^{|d_k|}}.
\label{prodexpansion}
\end{equation}
Deducting 1 and multiplying $A_k$ on both (\ref{nkexpansion}) and
(\ref{prodexpansion}), one is able to notice the 1-to-1 correspondence
between the terms of (\ref{nkexpansion}) and that of
(\ref{prodexpansion}). The correspondence reflects that the MLE can
only be achieved if the RHS (model) matches the LHS (data).

The above discussion unveils that if the LHS of (\ref{mgeneral})
matches the RHS term by term, the solution obtained from the equation
is the MLE. The model used by (\ref{mgeneral}) is based on the
assumption that every term of (\ref{nkexpansion}) exists. Note that
matching the RHS to the LHS, term by term, is also the condition that
(\ref{mgeneral}) holds. With the increase of probes, the variation of
the estimate decreases according to Fisher Information. This is also
reflected on the corresponding terms. As $n\rightarrow \infty$, one
can even use a single pair of the correspondences to form an explicit
estimator. For instance, the estimator proposed in \cite{DHPT06} is
based on the last pair of the correspondences. However, if $n<\infty$,
some terms on the LHS may not exist, and then we must consider:
\begin{enumerate} \item
 whether (\ref{mgeneral}) is still the likelihood equation of the
data  set? and \item whether the unique solution obtained from
(\ref{mgeneral}) is  the MLE?
\end{enumerate}

\subsection{No Solution because of Invalidity}
Recall lemma 1 of \cite{CDHT99} that states  there is a unique
solution to (\ref{mgeneral}) in the support of $A_k$ if $\sum_{j \in
d_k} \alpha_j
> 1$. This condition in practice means that there is at least an intersection in the observations of the descendants of node k, mathematically the condition can be
written as $\exists x, y; I(\{x,y\}) >0, x, y \in M\setminus
\emptyset$. Note that lemma 1 does not ensure the solution is the MLE
of $A_k$.

In contrast, if

\[
\sum_{j \in d_k} \alpha_j = \dfrac{\sum_{j \in d_k} n_j(1)}{n_k(1)}=
1,
\]
 there must have $\forall x,y; I(\{x, y\})=0,
x, y \in M\setminus \emptyset$. We call this complete mutual exclusion
at node $k$. Once the complete mutual exclusion occurs at node $k$,
there is no correlated information about $A_k$ in observation. Thus,
lemma 1 of \cite{CDHT99} concludes there is no solution to
(\ref{mgeneral}). This can be explained either by only considering
equation (\ref{mgeneral}) or by considering the validity of equation
(\ref{mgeneral}). \cite{CDHT99} takes the former and considers
(\ref{mgeneral}) a concave function that does not intersect with the
axis $A_k$ in $(0,1)$. We take the latter and consider if the complete
mutual exclusion occurs at node $k$, the loss rate of subtree $k$ is
equal to $1-\sum_{j \in d_k} \dfrac{\gamma_j}{A_k}$ instead of
$\prod_{j \in d_k}(1-\dfrac{\gamma_k}{A_k}\cdot \alpha_j)$. This means
that given the complete mutual exclusion, (\ref{mgeneral}) no longer
holds, let alone a solution.

\subsection{Incorrect Solution because of Invalidity}

As stated, if $n < \infty$ or $n \ll \infty$, some of the terms on the
LHS of (\ref{mgeneral}) may not exist, however, their counterparts on
the RHS do as long as $\hat \gamma_i \neq 0, i \in d_k $.  If so,
there is at least a mismatch between the LHS and the RHS. Then, the
unique solution obtained from (\ref{mgeneral}) may not be the MLE. We
call this partial mutual exclusion, mathematically
\[ \exists
x, I(x)=0; x \in M\setminus \emptyset.\] As the complete mutual
exclusion, (\ref{mgeneral}) does not hold if a partial mutual
exclusion occurs in observation. For instance, assume node $k$ has 3
descendants, a, b, and c, $I(\{a,b\})=n_{ab}(1)>0$, $I(\{a,c\})=0$,
$I(\{b,c\})=0$, and $I(\{a,b,c\})=0$. Putting the available
information into the expansion of (\ref{mgeneral}), we have
\begin{eqnarray}
\dfrac{n_{ab}(1)}{n}&=&\dfrac{n_a(1)n_b(1)}{n^2 \cdot A_k} +
\dfrac{n_a(1)n_c(1)}{n^2 \cdot A_k} + \dfrac{n_b(1)n_c(1)}{n^2 \cdot
A_k}- \nonumber \\
&&\dfrac{n_a(1)n_b(1)n_c(1)}{n^3\cdot A_k^2}. \label{inconsistent}
\end{eqnarray}
Although there is a unique solution to (\ref{inconsistent}), the
solution is certainly not the MLE of $A_k$ since the data and model
are obviously mismatched. In fact, the observation of $R(c)$ does not
provide any information for $A_k$ and should not be considered. If we
remove the terms related to subtree c, we have
\[ \dfrac{n_{ab}(1)}{n}=\dfrac{n_a(1)n_b(1)}{n^2
\cdot A_k} \] where the model fits the data.

As the previous subsection, (\ref{mgeneral}) is no longer valid if
there is a mutual exclusion in observation. Then, the following
questions are emerged:
\begin{enumerate}
\item how many types of exclusions exist in observation?
\item is there an estimator applicable to all types of observations?
\end{enumerate}

We will address those issues in the next section.

\section{Data Classification and Solutions}
\subsection{Classification of Data Set}

\begin{figure}
\centerline{\psfig{figure=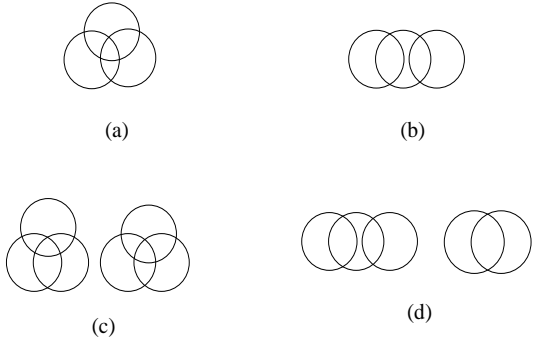,height=4.5cm,width=7cm}}
\caption{Data Classes } \label{class}
\end{figure}

The discussion presented in the last section unveils the impact of
observations on estimation and details the incompletion of previous
works on loss tomography. Considering the variation in observation, we
propose a new strategy in loss tomography to match a model to data. To
make this possible, we divide the data sets used in estimation into a
number of classes on the basis of intersection and partition in the
observations of descendants and introduces a number of models, one for
a class of data.  The classification is presented in Table
\ref{classtab}, where
\begin{itemize}
\item the perfect class denotes the data sets that satisfies the
follows:
\[
\forall x, I(x)>0, x \in M\setminus \emptyset;
\]
\item the chained-only is for the data sets that are not in the perfect class,
but the observations of the descendants cannot be divided into two
exclusive groups, i.e. $\exists x, y, I(\{x,y\})=0, x,y  \in M
\setminus \emptyset$ and
 \begin{eqnarray}
\mbox{if}&& (I(\{x, y\})=0, x, y \in M\setminus \emptyset) \mbox{
then} \nonumber \\ && \exists z, I(\{x,z\})>0\bigcap I(\{y,z\})>0, z
\in M; \nonumber
 \end{eqnarray}
\item in contrast to the chain-only, the partition only is for such
mutual exclusions that the observation of $R(k)$ can be divided into a
number of exclusive partitions and at least one partition has more
than 2 descendants. Within a multi-descendants partition, the
observation is perfect.
\item the chained partition class is for the data sets that combine
the characteristics of the above two classes, i.e. the observation can
be divided into a number of exclusive partitions and at least one
partition has more than 2 members, and in a multi-member partition,
its observation falls into the chained class.
\end{itemize}
Figure \ref{class} illustrates the four classes, where each circle is
for the observation of a descendant. Among the subfigures, (a) is for
the perfect class, (b) for the chained-only class, (c) for the
partition class, and (d) for the chained partition class.

\begin{table}
\centerline{\begin{tabular}{|l|r|r|}
  \hline
  % after \\: \hline or \cline{col1-col2} \cline{col3-col4} ...
  Observation Class  & partition & chain \\\hline
  perfect & 0 & 0 \\ \hline
                                                         chained-only & 0 & 1
                                                         \\ \hline
                                                         partition only& 1 & 0
                                                         \\ \hline
                                                         chained partition & 1 & 1 \\
  \hline
\end{tabular}}
  \label{classtab}
   \caption{Classification of Observations}
   \end{table}

\subsection{Prefect Observation}

As stated, most of the likelihood equations proposed previously do not
consider the variation of observations. With the introduction of data
classes, the likelihood equations proposed in the past need to be
calibrated to find their applicability in regard to the data classes.
The equations that are of concern in this paper is (\ref{mgeneral})
for the tree topology and (\ref{generalpoly}) for the general topology
because they are the most influential one in each of the topologies.
The following theorem provides the validity of (\ref{mgeneral}).

\begin{theorem} \label{perfectclass}
The estimate obtained from (\ref{mgeneral}) is the MLE iff the data
used in estimation falls into the perfect class.
\end{theorem}
\begin{IEEEproof}
The estimate obtained from (\ref{mgeneral}) has been proved to be the
MLE \cite{CDHT99}, where mutual exclusions in observations have not
been considered. On the basis of the analysis presented in the last
section,  the RHS of (\ref{mgeneral}) contains all possible terms of
correlations, from pairs of descendants to the product of all
descendants. This requires the LHS to have all correspondent terms to
match the RHS. Therefore, there must not have any form of mutual
exclusion in observation. On the other hand, if there is a mutual
exclusion in observation between the observations of siblings,
(\ref{mgeneral}) no longer holds. Then, the theorem follows.
\end{IEEEproof}

Another estimator is proposed recently  in \cite{ZD09} to tackle the
use of approximation to find the solution of (\ref{mgeneral}) if a
node has 5 or more descendants. \cite{ZD09} proposes
 an equivalent transformation to turn (\ref{mgeneral})
into a linear function by merging the descendants into two groups.
 The transformation takes advantage of
the self-similarity of (\ref{mgeneral}), and derives the follows:
\begin{eqnarray}
1-\dfrac{\gamma_k}{A_k}&=&\prod_{j \in d_k}(1-\dfrac{\gamma_k}{A_k}
\cdot \alpha_j) \nonumber \\
&=& \prod_{j \in k1}(1-\dfrac{\gamma_k}{A_k} \cdot \alpha_j) \prod_{j
\in k2}(1-\dfrac{\gamma_k}{A_k} \cdot \alpha_j) \nonumber \\
&=&(1-\dfrac{\gamma_{k1}}{A_k})(1-\dfrac{\gamma_{k2}}{A_k})
\label{group}
\end{eqnarray} where $k1$ and $k2$ denote the two groups which satisfy
$d_k=k1\bigcup k2$ and $k1\bigcap k2 = \emptyset$. Further, $k1$ and
$k2$ can be considered two virtual descendants of node $k$ and each
connects the descendants of its group. Note that the derivation of
(\ref{group}) takes advantage of theorem \ref{perfectclass}, where it
is assumed the observations of $k1$ and $k2$  to fall into the perfect
class. If $I(\{k1, k2\})>0$, it is easy to prove there is a unique
solution and the solution can be obtained analytically. On the other
hand, if there is a unique solution to (\ref{mgeneral}), the same
solution should be obtained from (\ref{group}) no matter how the two
groups are constructed. This requires $\forall x, I(x)>0, x \in
M\setminus \emptyset$. If $\exists x, I(x)=0, x \in M\setminus
\emptyset$, the $x$ can be selected as one of the two groups. However,
the $x$ itself is not in the perfect class. Thus, the assumption made
previously does not hold, and neither does (\ref{group}).
  The following corollary
provides the detail.

\begin{corollary}
If a data set belongs to the perfect class, the maximum likelihood
estimate can be obtained from (\ref{group}).
\end{corollary}

\subsection{Chained Only}

As stated, (\ref{mgeneral}) is only valid to the perfect data sets. If
the data set obtained from an experiment falls into the chained-only
class, a new likelihood equation is needed that can be obtained by
removing some of the terms from the RHS of (\ref{mgeneral}) that
correspond $I(x)=0, x \in M\setminus \emptyset$. Let $m_e, m_e \in M$
denote the set of the terms that need to be removed; if $j \in m_e$,
$|j|$ denotes the number of descendants involved in the term. Then,
the likelihood equation takes the following form
\begin{eqnarray}
1-\frac{\gamma_k}{A_k}=\prod_{j \in d_k}
(1-\dfrac{\gamma_k}{A_k}\alpha_j) - \sum_{j \in m_e} \big [
(-\dfrac{\gamma_k}{A_k})^{|j|} \prod_{l \in j} \alpha_l \big ]
\label{likelihood with exclusion}
\end{eqnarray}
where the summation on the RHS is for the terms that need to be
removed from the first term on the RHS.

Given (\ref{likelihood with exclusion}) as the likelihood equation,
the next question is whether there is a unique solution to it in the
support of $A_k$. The following lemma answers the question.

\begin{lemma} \label{unique solution}
Let C be the set of $c=(c_i)_{ i=1,2,...., d_k}$ with $c_i \in (0,1)$
and $\sum_i c_i >1$.  The equation of $1-x=\prod_{i \in d_k} (1-c_i x)
-\sum_{j \in m} (-x)^{|j|}\prod_{l \in j} c_l$ has a unique solution
$x(c) \in (0,1)$ if the summation term is a part of the product one.
Moreover, $x(c)$ is continuously differentiable on C.
\end{lemma}
\begin{IEEEproof}
See appendix
\end{IEEEproof}
The lemma extends from lemma 1 of \cite{CDHT99}, (\ref{likelihood with
exclusion}), as (\ref{mgeneral}), is a polynomial with a degree that
is lower than that of (\ref{mgeneral}) since at least $n_{d_k}(1)=0$.
Using the lemma, we can prove the solution
 to (\ref{likelihood with exclusion}) is the MLE.

If the chained observations of $\Omega_j, j \in d_k$ can be divided
into two exclusive groups, $k1$ and $k2$, where the observations of
$k1$ fall into the perfect class, and the observations of $k2$ are
exclusive partitioned, i.e. there is no intersection in the
observations of any two descendants of $k2$. In addition, the
observation of each descendant of  $k2$ is intersected with all of
$k1$, (\ref{group}) can be used here to obtain the MLE since . For
instance, if node $k$ has 3 descendants, $a$, $b$, and $c$, the
observations of the 3 descendants belong to the chained-only class,
where $I(\{a, b\})>0$, $I(\{a, c\})=0$, and $I(\{b,c\})>0$. If $a$ and
$c$ are in $k2$ and $b$ is in $k1$, we have the MLE from
(\ref{group}), where
\[
\dfrac{n_{ab}(1) + n_{bc}(1)}{n}=\dfrac{(n_a(1)+n_c(1))n_c(1)}{n^2A}.
\]
Hence, merging the descendants having exclusive observations into $k2$
is equal to remove those terms that do not occur in the data part from
the model part. Since each descendants in $k2$ is intersected with
every one of $k1$,  timing the statistic of $k2$ to that of $k1$
maintains the correspondences between data and model.

The above discussion shows that by appropriate grouping, (\ref{group})
can obtain the MLE for some of the data sets falling into chained
class. However, if a given data set cannot be divided into two groups
as above, the estimate obtained by (\ref{group}) is not the MLE.
Despite this, the estimate obtained by (\ref{group}) is still a little
better than that obtained by (\ref{mgeneral}). Using the previous
example and assume $I(\{a, b\})>0$, $I(\{a,c\})=0$, and
$I(\{b,c\})=0$. In this case, the observations cannot be merged into
two groups as above. If we merge b and c, and putting the statistics
into the expansion of (\ref{group}), we have
\begin{eqnarray}
\dfrac{n_{ab}(1)}{n}&=&\dfrac{n_a(1)n_b(1)}{n^2 \cdot A_k}
+\dfrac{n_a(1)n_c(1)}{n^2 \cdot A_k}. \label{inconsistent1}
\end{eqnarray}
As (\ref{inconsistent}), there is an inconsistence between the LHS and
the RHS of (\ref{inconsistent1}). Then, the estimate obtained from the
equation is not the MLE either despite that the error here is smaller
than that of (\ref{inconsistent}). On the other hand, if merging $a$
and $b$, we cannot even have a solution since (\ref{group}) fails to
hold.

\subsection{Partition-Only}
For the simplicity reason, the discussion is started from the
observation of $R(k)$ that consists of two exclusive partitions, and
then the discussion is extended to multiple partitions.

Let $k1$ and $k2$ be the two partitions. Thus, the LHS of the
likelihood equation is equal to 1 minus the sum of the pass rates of
$k1$ and $k2$ since their observations are mutual exclusive, and the
RHS of the equation according to (\ref{likelihood with exclusion}) is
equal to deducting those terms that involve the members of the two
partitions from the terms of the perfect class. The likelihood
equation is given in theorem \ref{2group partition}.

\begin{theorem} \label{2group partition}
For a network of the tree topology, if the observation of $R(k)$ is
partitioned into two exclusive parts, the likelihood equation of the
observation is as follows:
\begin{eqnarray}
2-\dfrac{\sum_{j \in \{1,2\}} \gamma_{kj}}{A_k}= \sum_{j\in
\{1,2\}}\big [\prod_{i \in kj}(1-\dfrac{\gamma_i}{A_k})\big ]
\label{likelihood with partition-only}
\end{eqnarray}
\end{theorem}
\begin{IEEEproof}
Let $ME$ denote the terms that need to be removed from
(\ref{mgeneral}) because of the mutual exclusion. Then, according to
(\ref{likelihood with exclusion}), we have
\begin{eqnarray}
1-\dfrac{\gamma_k}{A_k}&=&\prod_{j \in d_k}
(1-\dfrac{\gamma_j}{A_k}) - ME \nonumber \\
&=& \prod_{j \in k1} (1-\dfrac{\gamma_j}{A_k}) + \prod_{j \in k2}
(1-\dfrac{\gamma_j}{A_k}) -1 \nonumber \label{partition likelihood}
\end{eqnarray}
Removing $ME$ is equivalent to divide the descendants into two groups
according to the  mutual exclusion, and add the terms of one group
into another. Rearranging the terms of the above, we have
(\ref{likelihood with partition-only}).
\end{IEEEproof}

Given (\ref{likelihood with partition-only}), we have a polynomial
with the degree one less than that of the number of descendants in the
larger exclusive group. If the degree is larger or equal to 5, there
is no closed form solution to the polynomial unless some of the
descendants can be merged when we estimate the pass rate from the
source to their parent. Fortunately, this is achievable since the
observation of each group is perfect. Then, (\ref{group}) can be used
on each partition to turn (\ref{likelihood with partition-only}) into
a linear equation. Then, a closed form solution follows.

If the observation of node $k$, $\Omega_k$, is divided into Q
exclusive partitions (Q$>$2), we have the following theorem for its
likelihood equation.

\begin{theorem} \label{samesol}
If the observations of $R(k)$ are divided into Q partitions, the
likelihood equation is
\begin{eqnarray}
\sum_{j \in \{1,..,Q\}}(1-\dfrac{\gamma_{kj}}{A_k})= \sum_{j\in
\{1,..,Q\}}\big [\prod_{i \in kj}(1-\dfrac{\gamma_i}{A_k})\big ]
\label{multilikelihood with partition-only}
\end{eqnarray}
\end{theorem}
\begin{IEEEproof}
If the observation of $R(k)$ is divided into Q exclusive partitions,
it is equivalent to have Q independent subtrees connected to node $k$
and the pass rate of subtree $k$ is equal to the sum of the pass rates
of the Q independent subtrees. For each of the subtrees, there is a
likelihood equation as

\begin{eqnarray}
1-\dfrac{\gamma_{kj}}{A_k}&=&\prod_{j \in
d_{kj}}(1-\dfrac{\gamma_{kj}}{A_k} \cdot \alpha_j), \mbox{    } j \in
\{1,..,Q\}
 \label{dividing}
\end{eqnarray}
where $\hat\alpha_j=\dfrac{n_j(1)}{\sum_{j \in kj} n_j(1)}$.

Because of the independence in the observation of the Q partitions,
the likelihood
 equation is equal to the sum of the likelihood equations of
 the Q independent ones. Then, the theorem follows.

We can also prove (\ref{multilikelihood with partition-only}) from the
likelihood function constructed directly from the observation.
 Given the partitioned data set, the log-likelihood function of the observation in respect to $A_{k}$
 can
be written as
\begin{eqnarray}
L(A_{k}) =\sum_{j \in \{1,..,Q\}}\big [n_{kj}(1)\log A_{k}
+n_{kj}(0)\log(1-A_k\beta_{kj}) \big ] \nonumber
\end{eqnarray}
Differentiating it with respect to $A_{k}$ and letting the derivative
0, we have
\begin{eqnarray}
\dfrac{dL(A_{k})}{dA_{k}} &=& \sum_{j \in \{1,..,Q\}}\big
[\dfrac{n_{kj}(1)}{A_k}-\dfrac{n_{kj}(0)\beta_{kj}}{(1-A_k\beta_{kj})}
\big ]=0 \nonumber \label{derivative}
\end{eqnarray}
Solving it, we have
\begin{eqnarray}
\dfrac{\sum_{j \in \{1,..,Q\}}n_{kj}}{A_{k}}&=&\sum_{j \in
\{1,..,Q\}}\dfrac{n_{kj}(0)\beta_{kj}}{(1-A_k\beta_{kj})}
\nonumber \\
&=& n\cdot \sum_{j \in \{1,..,Q\}}\beta_{kj}  \nonumber
\end{eqnarray}
 Then, we have
\begin{equation}
A_k=\dfrac{\sum_{j \in \{1,..,Q\}}n_{kj}}{n\cdot \sum_{j \in
\{1,..,Q\}}\beta_{kj}}. \label{partsol}
\end{equation}
Note that if the observation of $kj$ falls into the perfect class, we
have
\begin{eqnarray}
\beta_{kj}&=&1-\prod_{j \in d_{kj}}(1-\dfrac{\gamma_{j}}{A_k}) \mbox{
,    }  j \in \{1,..,Q\} \nonumber
\end{eqnarray}
Using the above to substitute $\beta_{kj}$ from (\ref{partsol}) and
rearranging the terms afterwards, we have (\ref{multilikelihood with
partition-only}).

\end{IEEEproof}

Since (\ref{dividing}) is a concave function, (\ref{multilikelihood
with partition-only}) is a concave function because
(\ref{multilikelihood with partition-only}) is a sum of
(\ref{dividing}). In addition, there is a common support for each of
the member
 functions of (\ref{multilikelihood with partition-only}). Then, (\ref{multilikelihood with partition-only}) has a maximum point in
 $(0,1)$ that can be obtained directly by

 \[
 \hat A_k=\dfrac{\sum_{i \in \{1,..,Q\} }\hat \gamma_{k_{i1}}\hat
 \gamma_{k_{i2}}}{\sum_{i \in \{1,..,Q\}} n_{ki}(1)}
 \]
where  (\ref{group}) is used in each of the exclusive partitions to
divide the descendants into two groups and merge their statistics,
where $n_{ki}(1)$,  $i \in \{1,2\}$, is the number of probes that
observed simultaneously by the receivers of the two groups of the
$ith$ exclusive partion. In addition, $\hat \gamma_{k_{i1}}$ and $\hat
 \gamma_{k_{i2}}$ are the empirical pass rates from the source to the
 two groups of the $ith$ exclusive partition, respectively.

\subsection{Chained Partition}
As defined, a data set falling in this class can be divided into a
number of exclusive partitions, each partition consists of the
observation of a number of descendants. In addition, the observation
of a partition that has more than 2 members is not in the perfect
class but chained. Thus, a new likelihood equations is needed for this
class of data and the equation must combine the feature of the
likelihood equations proposed for the chained-only and partition only
classes. Since the observation of a partition is not in the perfect
class, the likelihood equation for the partition is in the form of
(\ref{likelihood with exclusion}). If there are Q ($Q>1$) partitions,
since the observations between partitions are exclusive, the
likelihood equation for this class is in the form of

\begin{eqnarray}
Q-\dfrac{\sum_{j \in \{1,..,Q\}} \gamma_{kj}}{A_k}&=& \sum_{j\in
\{1,..,Q\}}\big [\prod_{i \in kj}(1-\dfrac{\gamma_i}{A_k})- \nonumber
\\
&&\sum_{ p\in m_e(j)} (-\dfrac{\gamma_k}{A_k})^{|p|} \prod_{l \in p}
\alpha_l \big ]. \label{multilikelihood with chained}
\end{eqnarray}
As before, we can prove there is a unique solution to
(\ref{multilikelihood with chained}) since (\ref{multilikelihood with
chained}) is a sum of concave functions. Unfortunately, there is no a
closed form solution to (\ref{multilikelihood with chained}) at this
moment.

\subsection{Complete Mutual Exclusion}

Given Theorem {\ref{samesol}}, we know if there is no intersection
between the observations of the receivers attached to the subtrees
rooted at node $k$, the observations of the descendants do not provide
any information about the path connecting the source to their parent.
Thus, there is no need to add the correlation between them into the
model part of (\ref{mgeneral}). The RHS of (\ref{mgeneral}) is equal
to
\begin{equation}
1-\sum_{j \in d_k} \dfrac{\gamma_j}{A_k}. \label{equal} \end{equation}
Using empirical probability $\hat \gamma_j=\dfrac{n_j(1)}{n}$ to
replace $\gamma_j$ from (\ref{equal}), (\ref{mgeneral}) is collapsed
and there is no solution of course.

To solve the problem,  we can either sending more probes to break the
tie or use bootstraps to produce some synthetical probes to create
intersections. Then, we can use an appropriate likelihood equation to
estimate $A_k$.

\subsection{Independent Path}
 During the presentation, we always assume that either there are at least 2
 descendants in a partition or at least there is a partition that has 2
 descendants of node $k$. Without this assumption, a data set may be
 in the class of complete mutual exclusion. Under this assumption, if
 each partition has more than 2 descendants, the pass rate from the source to the parent of the descendants can
 be estimated independently and the total pass rate is equal to the
 sum of the pass rates of the partitions.

Although using Theorem \ref{samesol}, we are able to obtain the same
solution as that obtained from (\ref{mgeneral}) and revive the
estimator based on the equivalent transformation. A new question is
emerged that is whether an estimator needs to consider a partition
that only has a single descendant.
 The following theorem provides the answer to this question.

 \begin{theorem}\label{independent path}
 The observation of single-descendant groups  has no impact
 on the estimation of the pass rate of the path connection the source
 to the parent of the descendant.
 \end{theorem}
 \begin{IEEEproof}
 According to the condition, (\ref{multilikelihood with partition-only})  is the likelihood equation fitting the data.
For this case, a number of identical terms can occur in the summation
of the LHS and RHS of the equation, one for the loss rate of a
single-descendant group. Those terms cancel each other without effect
to the estimation. Then, the theorem follows.
 \end{IEEEproof}
For instance, in the previous example descendant $c$ is independent
from $a$ and $b$. Based on theorem \ref{independent path}, c is
removed from estimation and we have
\begin{equation} \label{lemma2 example}
\dfrac{n_{ab}(1)}{n}=\dfrac{n_a(1)n_b(1)}{n^2 A_k}.
\end{equation}
Using lemma \ref{2group partition}, we have
\begin{eqnarray}
&& 2-\dfrac{n_a(1)+n_b(1)-n_{ab}(1)+n_c(1)}{nA_k}= \nonumber
\\
&& \mbox{   } (1-\dfrac{n_a(1)}{nA_k})(1-\dfrac{n_b(1)}{nA_k}) +
(1-\dfrac{n_c(1)}{nA_k}). \nonumber
\end{eqnarray}
Simplifying the above, we have (\ref{lemma2 example}). Theorem
\ref{independent path} also explains why there is no solution for the
data set falling into the complete mutual exclusive group. This is
because the observation of each descendant has its own partition, and
then each of them is canceled from (\ref{multilikelihood with
partition-only}) that leads to the collapsed of (\ref{multilikelihood
with partition-only}).

\section{Multi-sources}
In the general topology, a node may have more than one parents, thus a
node may observe probes sent by multiple sources. Because of this,
estimating the pass rate of a link must consider the probes sent from
all related sources regardless the probes pass the link of interest or
only pass its ancestors.  Therefore, (\ref{generalpoly}) is the
likelihood equation of $A(s,i)$ for a path connecting source $s$ to
node $i$ regardless node $i$ is a joint node or not, where a joint
node is such a node that has more than one parents \cite{Zhu09}. The
difference between (\ref{generalpoly}) and (\ref{mgeneral}) is at
those nodes that can receive probes from multiple sources; where the
former considers all probes sent by related sources to node $f(j)$
 and uses
\begin{equation} \label{multisource alpha} \hat\alpha_j=\dfrac{\sum_{s
\in S(i)} n_j(s, 1)}{\sum_{s \in S(i)} n_{f(j)}(s,1)}
\end{equation}
as the empirical pass rate of link j, while the latter has
\[
\hat\alpha_j=\dfrac{n_j(1)}{n_{f(j)}(1)}.
\]
Note that in (\ref{multisource alpha}) $i$ is the parent node of link
$j$ and $S(i)$ is the set of sources sending probes to node $i$.
Despite the differences between the likelihood equations of the tree
and the general topologies, both take into account all probes reaching
the end node of the path of interest. More importantly, the principle
of fitting a model to data becomes more obvious in the general
topology than that in the tree topology. If we consider the both sides
of (\ref{generalpoly}) data and model, respectively, the LHS of
(\ref{generalpoly}) as the data is from the observation of a single
source, called individual observation; while the RHS as the model
considers the probes sent by multiple sources, called global
observation. As previously proved, the MLE can be obtained if and only
if the RHS matches the LHS and there is a unique solution to
(\ref{generalpoly}).

Considering fitting a model to data in the general topology, we can
use the same classification as those defined in the tree topology to
divide data sets into 4 classes. Since the nodes in a general topology
can have multiple parents, even a node that has only a parent can have
multiple ancestors located on different paths to the node, the nodes
are divided into 3 types: single parent and single source nodes
(single node), multiple parents nodes (joint nodes), and single parent
and multiple sources nodes (shared nodes). For the single parent and
single source nodes, the methods proposed for the tree topology can be
used to estimate the loss rate of the path connecting the source to
the node given the loss rates of the subtrees rooted from the node, in
particular if there are shared segments in the subtree. Our focus is
on the joint nodes because single nodes need to know the pass rate of
the shared subtrees and a shared node can be considered a special
joint node. The difference between a joint node and a shared one is on
the paths connecting the sources to the node, where the former has
distinguished paths and the latter has a shared part of the paths.
Therefore, they can be handled in the same manner in terms of
estimation. In addition, if we use the divide-and-conquer approach
proposed in \cite{ZD09} to divide a general topology into a number of
trees, there is no need to consider the shared nodes separately.

\subsection{Joint Node}

For a joint node, say $i$, there are up to $|S(i)|$ likelihood
equations, one for a path connecting a source to the node.
\begin{table}
\centerline{\begin{tabular}{|l|r|}
  \hline
  % after \\: \hline or \cline{col1-col2} \cline{col3-col4} ...
  {\bf Individual Obs.} & {\bf Global Obs.}\\\hline
  Prefect & Prefect \\ \hline
  Others & Prefect                        \\ \hline
Others & Identical Others \\ \hline
  Others & Others \\
  \hline
\end{tabular}}
  \label{multi-obstab}
   \caption{Observation Classification}
   \end{table}
Since all the paths connect to an ordinary subtree rooted at node $i$,
we need to have a unique pass rate for the subtree that can maximize
the likelihood function constructed from observation. As previously
analyzed, each of the likelihood equations is determined by
observation. (\ref{generalpoly}) as (\ref{mgeneral}) holds if and only
if observation is in the perfect class, i.e. $\forall j,
\Omega_i(s_j), s_j \in S(i)$ is in the perfect class. In this case,
the $|S(i)|$ likelihood equations are equivalent to each other in
regard to the shared subtree and the LHS of (\ref{generalpoly})
matches its RHS. Knowing the pass rate of one of the paths, say from
$s_k$, the pass rate of another path, say from $s_j$, can be obtained
easily by

\[
A(S_j,i)=\dfrac{\hat \gamma_{S_j}}{\hat \gamma_{S_k}}A(S_k,i)
\]
since the $|S(i)|$ likelihood equations are in the form of
$1-x=\prod_{j \in d_i}(1-c_jx)$. Given the fact that subtree $i$ is a
common part of the paths from the sources of $S(i)$ to $R(i)$, the
pass rates of the paths from the sources to node $i$ are proportional
to the pass rates from the source to $R(i)$.

If $\exists \Omega_i(S_j), S_j \in S(i)$ is not in the perfect class,
the situation becomes more complex than that of the tree topology and
needs to be analyzed further. To assist the following discussion, we
divide the observations of a shared subtree into 4 classes on the
basis of individual and global observation and present them in Table
\ref{multi-obstab}. The global observation of node $i$ is defined as
\[
\Omega_i=\bigcup_{s_j \in S(i)} \Omega_i(s_j). \] Apart from the {\it
(perfect, perfect)} class, we need to consider data consistency again
in a different way from those defined in \cite{CDHT99} and
\cite{ZD09}. The previous concern is focused on the consistency
between data and model and the approach used is to eliminate those
data sets that is inconsistent with the model. Here, the consistency
has been extended to consider the consistency between individual
observations, and the consistency between an individual observation
and a global one. With multiple sources sending probes to receivers,
each source creates its own individual observation that is the view of
the source on the shared segment. The views created by different
sources can be different from each other although when $n^s
\rightarrow \infty, s \in S(i)$ the same views are expected. However,
when $n^s < \infty$, different views may occur that make estimation
impossible since there is a lack of a consistent model.

For the data set in the {\it (others, perfect)} class, although the
data created by different sources compensate each other to create a
perfect view, the individual data are not consistent with the global
one that implies different models for the likelihood equations.
Because of this, estimation cannot proceed and we need to send more
probes until the data falls into {\it (perfect, perfect)} class or
skip the estimation. This also apply to the data of {\it (others,
others)} class since if $\Omega_i(S_i)$ is not consistent with
$\Omega_i(S_j)$, the $|S(i)|$ likelihood equations are different from
one another.

The {\it (others, identical others )} class is designated to the
observation: $\forall \Omega_i(s_j), s_j \in S(i)$ are identical in
terms of correlation. Hence, if the individual observations are
identical to the global one in terms of correlation, a model that is
consistent with the data can be created, so does a consistent
likelihood equation for every source. This extends (\ref{generalpoly})
 to cover imperfect data in some degree and the following theorem presents the likelihood equation for
the pass rate of a shared subtree.
\begin{theorem}
Given data in the {\it (others, identical others )} class, the
likelihood equation of the pass rate of the shared subtree is as
follows:
\begin{equation}
|S(i)|(1-x)=|S(i)|\prod_{j \in d_i}(1-\alpha_j x) - \sum_{j \in d_i}
me_{j}(x) \label{multi-inperfect}
\end{equation}
where $\alpha_j$ is the ratio between the number of probes reaching
$j, j \in d_i$ and the number of probes reaching node $i$  as
(\ref{multisource alpha}) and $me_j(x)$ as defined in the tree
topology, denotes the terms that need to be removed from likelihood
equation $j$ (see proof for detail).
\end{theorem}

\begin{IEEEproof}
Let $x=\dfrac{\gamma_j}{A(s_j,i)}$ that is the pass rate of the shared
subtree. If $\Omega_i$ is not perfect, the corresponding terms on the
RHS of (\ref{generalpoly}) should be removed as those discussed in the
tree topology. Let the terms be $me_j(x)$ for $s_j$ that is a function
of $x$. We have a likelihood equation as follows for each source:
\[
(1-x)=\prod_{j \in d_i}(1-\alpha_j x)-me_j(x).
\]
There are $|S(i)|$ equations as above, one for a source. Adding the
equations together, we have the theorem.
\end{IEEEproof}
As previous, we are able to prove the solution space is concave and
there is a unique solution in the support of the pass rate. Then, the
solution is the MLE of the pass rate of the shared subtree.

 Given the
pass rate $x$, the loss rate of the path connecting a source to node
$i$ can be obtained directly by $\dfrac{\hat \gamma_j}{x}$ since
(\ref{multi-inperfect}) is a polynomial, if its degree is 5 or higher,
there is a lack of explicit methods so far to solve
(\ref{multi-inperfect}) other than approximation. To minimize the use
of approximation,
 we can use the divide-and-conquer approach proposed in
\cite{Zhu09} here to break a general network into a number of trees,
where (\ref{generalpoly}) or (\ref{multi-inperfect}) is used on each
joint node to have the MLEs of the number of probes reaching the joint
node. With the numbers, a general network can be divided into a number
of trees and the methods proposed in the previous section can be used
to obtain the MLE of each path.

\section{Conclusion}
Loss tomography is built on statistical inference that requires a
correct model to describe the observation received from an experiment.
The model must match the nature of the data. Nevertheless, the
dependency of a model on a set of data has been either overlooked or
misunderstood that leads to a misconception that an estimator is
applicable to all sorts of data sets as far as it returns a unique
solution. Within this paper we attempt to correct this and consider
the validity of an estimator that demands a match between data and
model in estimation. We then revisit two of the most influential
estimators proposed previously and find that they, as those estimators
proposed previously, at most are the maximum likelihood estimator for
a type of data only.  To overcome this, fitting a model to data has
been emphasized in this paper and the necessary and sufficient
condition of the maximum likelihood estimator is presented in this
paper that require 1) a likelihood equation matching a model to the
data; and 2) a unique solution to the likelihood equation. The
necessary and sufficient conditions indicates that in order to obtain
a MLE, we need to use all available information in observation,
eliminate redundant information, and match a model to the data. To
generalize the results, data obtained from experiments are divided
into 4 classes, and 4 likelihood equations are presented in this
paper, one for a data class. Apart from the tree topology, this issue
is also considered for the general topology, where data consistency
has been extended to consider the difference between individual views
and between an individual view and the global view. The connection and
relation between them have been analyzed and an estimator is proposed
for the case of identical individual views.

\section*{Appendix}

\noindent{\bf Lemma \ref{unique solution}}
\begin{IEEEproof} Let $h_1(x)=1-x$, $h_2(x)=\prod_i (1-c_i x)$, and $h_3(x)=\sum_{j \in m_e} x^{|j|} \prod_{e \in j} c_e$, we
have $h_1'(x)=-1$, $h_2'(x)=h_2(x)\sum_i\frac{c_i}{(1-c_i x)}$, and
$h_3'(x)= \sum_{j \in m_e} |j|*x^{|j|-1} \prod_{e \in j} c_e$. Let
$q_i= \sum_i\frac{c_i}{(1-c_i x)}$, we have $h_1^{''}(x)=0$,
$h_2^{''}(x)=h_2(x)[(\sum_i q_i)^2 - \sum_i q_i^2]>0$, and
$h_3^{''}(x)= \sum_{j \in m_e} |j|*(|j|-1)x^{|j|-1} \prod_{e \in j}
c_e<0$, if $x \in [0,1]$. Note that $h_3(x)$ is a small part of
$h_2(x)$ that have two or more $c_i, i \in d_k$ timed together. Let
$h(x)=h_1(x)-h_2(x)+h_3(x)$, that is strictly concave on $[0,1]$.
 \end{IEEEproof}

\bibliography{congestion}

\end{document}